# Venturing the Definition of Green Energy Transition:
## A systematic literature review

Pedro V. Hernández Serrano[a] Amrapali Zaveri[a]
[a] *Institute of Data Science at Maastricht University,*
*{p.hernandezserrano; amrapali.zaveri }@maastrichtuniversity.nl*

**Abstract.** The issue of climate change has become increasingly noteworthy in the past years, the transition towards a renewable energy system is a priority in the transition to a sustainable society. In this document, we explore the definition of green energy transition, how it is reached, and what are the driven factors to achieve it. To answer that firstly, we have conducted a literature review discovering definitions from different disciplines, secondly, gathering the key factors that are drivers for energy transition, finally, an analysis of the factors is conducted within the context of European Union data. Preliminary results have shown that household net income and governmental legal actions related to environmental issues are potential candidates to predict energy transition within countries. With this research we intend to spark new research directions in order to get a common social and scientific understanding of green energy transition.

## 1. Introduction

The issue of climate change has become increasingly salient in the past years and the transition to a sustainable society is the main challenge of this century. The energy sector is the source of two thirds of global greenhouse gas emissions (IRENA, 2018) and the traditional power generation based on fossil fuels causes both environmental and health problems (McMichael, Montgomery & Costello, 2012). Therefore, the transition towards a renewable energy system is a priority in the transition to a sustainable society. In order to reach a long-term and cost-effective energy transition that addresses the economic and political complexities related to such structural transformation of not only the energy sector but of society as a whole, it is of primary importance that policymakers and stakeholders understand what is energy transition and what are its underlying drivers.
Renewable energy is at the core of the European Union's (EU) priorities, and the EU is currently working towards its 2020 targets and deployed 17.52% of renewable energy in gross final energy consumption in 2017 (European Commission, 2019b). However, the pace of increase of the renewable energy share has slowed down in the past years (European Commission, 2019B), and much is yet to be done for Member States to reach their more recently set 2030 renewable energy targets.
Although energy transition is today a priority, there is no consensus on the definition of energy transition. While some understand it as a transformation of the energy sector, others understand it as a broad transformation of society as a whole. In the academic sphere, many different factors and perspectives of energy transition are researched, suggesting that in reality energy transition might not only be the replacement of fossil fuels with renewable energy, but a comprehensive change in society regarding how energy is perceived. Factor is here understood as a measurable fact that allegedly contributes to a result. For example, some authors look at governmental action and institutional change (Bjørner & Jensen, 2002; Ruszel & al., 2017), taking a policy analysis perspective, while other studies focus on the energy sector and economic aspects (Sattich, 2014), taking an economic perspective. Moreover, in a majority of the literature, there is a lack of empirical analysis to support theoretical conclusions.

## 2. Approach

This research aims to contribute to the research on energy transition. Firstly, by asking what energy transition is, as an attempt to shed light on the lack of consensus regarding the definition of energy transition. Secondly, by asking what the different factors are responsible for energy transition. To answer these questions, key factors are derived from the academic literature and reports through a systematic literature review. Those key factors are then tested through a preliminary analysis in an attempt to assess how responsible they are regarding energy transition. By combining qualitative and quantitative analyses, our research contributes in a more relevant manner to the research on energy transition by investigating a variety of factors mentioned in the academic literature empirically. Our research identifies four recurrent themes of factors understood as responsible for energy transition by the academic literature: government action, market factors, public and technological advancement. Additionally, our research finds that accessible, relevant data is not easily found. Although data concerning renewable energy is available, there is no available data concerning the factors responsible for energy transition. When our research attempted to find data to test for factors mentioned in the literature, it was hindered by the lack of data focused on certain factors. Consequently, more accessible, machine-readable data regarding factors responsible for energy transition is crucial for a better understanding of what is driving energy transition. Our research adds a new multi-disciplinary work together with empirical investigation on the influence of different factors on energy transition. The focus of our study is the EU and its neighboring countries, as its specific stance as policymakers and leaders in the climate change movement provides plausible reasons to look into the specific factors influencing the energy transition in Europe. In the following Section, the literature review on energy transition is presented. Section 3 and 4 respectively present the case study of the EU and the methodology of this research. Moreover, the data used in the quantitative analysis is described. The preliminary results are presented and interpreted in Sections 5 and 6. Section 7 discusses the implications and limitations of the study and proposes complementary research.

## 3. Different Perspectives of Green Energy Transition

When reading about energy transition in the academic literature one can quickly notice the many different perspectives taken to study the topic. It is approached from different academic disciplines as well as in multidisciplinary work. Our research attempts to present a cross-discipline overview of the topic. This literature review is organized into four recurrent themes identified in the academic literature, namely the role of the government, of the market, of the public and of technological advancements. Although our research distinguishes between these themes, the later are very inter-connected and thus overlap. Additionally, this section discusses the relevance of reports on energy transition. Those mostly focus on presenting data and are therefore considered as primary sources. Moreover, this section presents certain studies that stand out from the rest.

A narrow definition is the transition from one state of an energy system to another, currently from using non-renewable energy sources, such as fossil fuels, to an energy system mostly based on renewable sources (Cleveland & Morris, 2006). Although most authors agree on this narrow definition, they still use a variety of different definition. Sattich (2014) describes energy transition as "substantially increasing the share of renewables in the energy mix, while phasing out nuclear power" (p.264), while Sung and Park (2018) define energy transition as a "collective, complex and long-term process comprising multiple actors for social changes, involving far-reaching societal changes" (p.2). In some cases, the economic aspect is more dominant, such as when defining energy transition as a transition of the economy into a low emission economy (Ruszel, Młynarski & Szurlej, 2017). Meadowcraft (2009) identifies this lack of consensus and puts forward that it is often unclear exactly which system is at interest and what sort of transition is on the table. He establishes over six different definitions with multiple possible variations, concluding that each type of transition implies a different mix of energy technologies and orientation for policy intervention. Lastly, another relevant question that presents many disagreements is whether nuclear energy is considered a low emissions energy source or not regarding its environmental impact (Meadowcraft 2009). As this question remains unanswered and is more political, nuclear energy is not considered as green energy in this research.

Our research chooses to use a broad definition of energy transition as it attempts to build a broad and comprehensive understanding of the topic. Hence, the definition by Sung and Park (2018) is used as it correctly acknowledges the complexity of the phenomenon, which is what this study attempts to highlight. Consequently, energy transition is understood here as a "collective, complex and long-term process comprising multiple actors for social changes, involving far-reaching societal changes" (Sung & Park, 2018, p.2)

## 3.1 Government action

A recurrent theme in the energy transition literature is the role of government actions and policies. Here government is understood as the public institutions of a state.

Some authors analyze the role of the government in energy transition by assessing and comparing government's policies (Rosenow & Kern, 2017; Lund, 2007; Bjørner & Jensen, 2002; Ruszel & al., 2017; Woerter, Stucki, Arvanitis, Rammer, & Peneder, 2017; Sung & Park, 2018). The policy analysis literature on energy transition attempts to assess the effectiveness of policies by looking at their impact on energy production and consumption in single case settings or by comparing several countries. Ruszel et al. (2017) edited a comprehensive book presenting a comparison of energy policy transition in European countries and the United States. Some authors also focus on the energy policy evolution in the EU (Rosenow & Kern, 2017; Fouquet & Johansson, 2008). They focus on the process of this evolution in the wider European integration context and consider the difference in implementation of these policies in the Member States.

Although regulations play a big role in energy transition, the government can also influence the transition by concluding voluntary agreements with energy-extensive companies. These are different arrangements between the government and businesses, where for example the company commits to carrying out certain energy saving activities and in return qualifies for a reduced energy tax rate (Bjørner and Jensen, 2002). Voluntary agreements are often backed up by the threat of alternative regulation and are sometimes found to be more effective than regulations themselves.

In the legal academic literature, energy transition is studied indirectly mostly by looking at the number of laws passed in the field of energy law and environmental law. The research is thus almost exclusively quantitative or focused on what drives law-making in those fields, but not on the quality of existing energy legislation (Fankhauser, Gennaioli & Collins, 2016). Fankhauser et al. (2016) acknowledge this gap and call attention to the fact that more laws do not necessarily correlate with stronger climate policy. However, a few articles analyze isolated cases of possible impact of a law on energy transition. For example, Dreyfus and Allemande (2018) analyzed how the French flagship law had an impact on energy transition in France. Law undoubtedly plays a role in any development of society. Zillman, Roggenkamp, Paddock and Godden (2018) even argue that energy transition consists of a mixture of technological and legal innovations. Although the pattern is far from uniform and there is no assessment of the impact of law-making on energy transition, the legal academic literature emphasizes the legislative role of the government in energy transition.

This focus on government actions, both executives and legislatives, in the literature on energy transition reflects the relevant role of the government in the process of energy transition.

### 3.2 The market

The market plays an important role in the transition to a green energy economy and several studies can be found on this topic. In our research, the market is identified as all the economic factors that can be reconnected to green energy such as subsidies, taxes, production, prices and income level.

Many authors found that investment in green energy, have a strong impact on energy transition. For example, Midilli, Dincer and Ay (2006), combined different measures of private and public financial support to study the total green energy budget of a country. The study finds that investment in green energy supply and progress can make an important contribution to the economies of the countries where green energy is produced. Moreover, some emphasize the role of disposable income in explaining why countries are in different stage of the transition; following the theory that consumption of renewable energy requires a premium to be paid and that in many advanced countries, fossil fuel is highly subsidized (Ruszel & al., 2017). Regarding energy prices, which affect the transition, Sung and Park (2018) found that the traditional energy industry has a direct negative impact on energy transition, given by the extensive propaganda concentrated on cheaper fossil fuel relative to renewable energy, which exploits the society's preference for affordable prices. They even identify an actor called "market" comprised of bank groups, energy economy, renewable energy platforms, firms and energy finance. This confirms the importance given to the market by the literature.

Other authors, like McCarl and Hertel (2018), analyzed the transition taking a climate approach. They find that climate change could have positive impacts on agriculture and therefore on the economy of some regions and negative in others.

Considering the importance given to prices, investments and incomes regarding the transition, the market appears as a recurrent theme in explaining energy transition.

### 3.3 The public

Factors relating to the public, encompassing phenomenon relating to citizens, are also present in the academic literature relating to energy transition. However, their exact importance is highly debated. Factors relating to the public seem to be overlooked more often than others, while some studies consider the public to be a crucial actor (Tsagarakis & al., 2018; Zahari & Esa, 2016).

The level of environmental concern, public awareness and information are seen as significantly influential regarding green energy transition (Ruszel & al., 2017; Midilli & al., 2006; Zillman & al., 2018). Some authors consider the public as vital for development of renewable energy (Zahari & Esa, 2016) and even consider the public sphere to have more influence on energy transition than the government itself (Meadowcroft, 2009). Additionally, education is discussed as having an independent influence on the adoption of renewable energies (Zahari & Esa, 2016; Tsagarakis & al., 2018; Pachauri & Jiang, 2008; Sardianou & Genoudi, 2013). The actions of NGOs as well as public movements against government actions regarding energy are ways in which the public's opinion can be influential. Citizens in their role of consumers is a separate topic of discussion, and whether to consider it as a part of market related factors or one of civil society is unclear. The public sphere can also be seen as having only an indirect influence on energy transition, as was concluded by Sung and Park (2018). Although considered important, the study claims that the public does not seem influential enough to lead to social innovations directly contributing to the status and development of renewable energy system. The study however saw some positive causal relationships between the public, the government, the market actors and energy transition. This suggests that the public contributes to the transition by generating discussion and mobilizing social innovation, thus indirectly having an impact.

Income as well as different age groups are also considered to be relevant in certain studies relating to energy transition, mostly as positively affecting citizens attitude towards renewable energy projects (Sardianou & Genoudi, 2013). Pachauri and Jiang (2008) found that the income level of a certain region was seen as having a clear positive impact on the speed of energy transition. However, it is relevant to mention that the energy transition analyzed was the transition from low efficiency solid fuels to modern energy sources, such as electricity, in India and China. Hence, it is unsure that this conclusion applies in the energy transition currently taking place in Europe.

It is debatable whether the public has a direct or only an indirect effect on energy transition. One can however conclude that in any case, it is considered to influence it. Therefore, it is considered as a recurrent theme in the literature, impacting energy transition.

### 3.4 Technological advancement

Technological development in the energy sphere is moving rapidly and is considered essential for energy transition in a majority of the literature (Midilli & al., 2006; Pachauri & Jiang, 2008; EEA, 2017; Zillman & al., 2018). Some authors and most reports even present it as the most important factor (Rosenow & Kern, 2017). Some of the most important aspects seem to be the development of new forms of energy sources and the availability and productivity of the existing technologies (Midilli & al., 2006). Moreover, innovation at the level of energy systems, such as correct energy infrastructure to match power production with consumption, is also considered important (Sattich, 2014; Ruszel & al., 2017). Additionally, a stable and secure supply of energy is often brought up as enhancing energy transition. An example of this is the development of grid energy storage, different methods used to store electrical energy within an electrical power grid, to help balance energy supply and demand and allows for more large-scale manufacturing of energy (Ruszel & al., 2017; Pachauri and Jiang, 2008).

One technical factor which seems to be especially relevant is energy efficiency by optimum use of energy resources, described as the most cost-effective way of reducing gas emissions (Ruszel & al., 2017; IRENA, 2018; Zillman & al., 2018). Energy efficiency is considered equally important next to other technological advancements such as renewable energy technologies (Rosenow & Kern, 2017). It is even one of the three key pillars identified in the EU 2020 Strategy regarding climate change and energy, which includes a 20% reduction of projected primary energy consumption by 2020.

However, while the emergence of new technologies may be an essential component in energy transition, it is rarely considered sufficient in itself. Hence, the correct interaction with legal framework, public policy, research and development and investment remains crucial, as recognized by some of the literature (Zillman & al., 2018; Rosenow & Kern, 2017). Furthermore, energy price drops will also affect the speed of their adoption (Ruszel & al., 2017; Pachauri and Jiang, 2008). Due to the frequent mentioning of technological advancement in the literature, it seems the theme is given considerable importance as a factor. It is however also recognized that energy transition requires more than just inventing new technologies.

## 3.5 Geography

Finally, a group of other miscellaneous factors, mostly relating to geography and the political structure of a region, remain important but do not fit within any of the above-mentioned categories. A region's geographical location, energy balance structure and potential to develop renewable energy sources by having the necessary environmental conditions are considered having an impact on the speed and reach of energy transition (Kurtyka, 2017; Ruszel & al., 2017). Additionally, the political climate and questions of energy security can also change the course or speed of a region's energy transition. Moreover, international cooperation is sometimes brought up as an important contributor to energy transition, and even climate change as such is considered a catalyst for energy transition (Ruszel & al., 2017; Zillman & al., 2018; Fankhauser & Jotzo, 2018).

## 4. Related Work and Official Reports

Although most articles focus on one of the themes discussed, some articles take a multi-dimensional perspective and offer useful insights regarding energy transition. Sung and Park (2018) explore how various actors influence the transition to a renewable-energy economy. The study examines how four actors influence this transition: government, public, markets and the traditional energy industry. The authors conduct a text-mining analysis of the academic literature and tests the influence of actors on the transition to a renewable-energy economy through an empirical analysis. The authors find that the government and market actors have a positive impact on the transition while the traditional energy sector affects the transition negatively. While the authors find that the public has no direct influence on the transition, they find that the public and the government have positive indirect influence on energy transition by interacting with the market (Sung & Park, 2018).

Green energy transition literature is also represented on reports published by different institutes and organizations. These include a much more fact-based discussion than the rest of the literature on the state of green energy and energy transition in different regions. Reports discuss the benefits of energy transition but focus less on how to transition. Several reports (World Bank, 2018; EEA, 2017; Energy Atlas, 2018; European Commission, 2019b; IEA & the World Bank, 2017; IEA, 2017; IRENA, 2018) describe the current state of renewable energy in certain regions or globally, stating which energy sectors would need greater efforts, as well as the costs and consequences of energy transition. However, these do not mention how progress could be achieved.

The International Energy Agency (IEA), sometimes considered the world's most influential source of energy information (Oil Change International & IEEFA, 2018), publishes extensive yearly reports open to the public by purchase, outlining the development of the global energy system. However, the five pages executive summary of the 2017 report (IEA, 2017), available to the public for free, gives a glimpse into the state of the renewable energy sector and concludes with key recommendations for policy makers. They outline that the necessary change is happening in the energy system, but that policy signals are needed for acceleration of the energy transition. The IEA argues that the emphasis should be put on developing a long-term energy vision, enhancing international collaboration, supporting technology innovation as well as new business models and finally, increasing policy makers' understanding of digitalization in the energy sector. It is however worth noting that the IEA has been criticized for consistently making inaccurate predictions of energy transition, drawing up misleading energy targets (exceeding the limits set out by the Paris Agreement) and being too closely connected to the fossil fuel industry, suggesting that IEA might actually be hindering governments from achieving energy transition and climate change goals (Oil Change International & IEEFA, 2018).

The International Renewable Energy Agency (IRENA) also published a report in 2018 (IRENA, 2018). The later focuses more on data and on the energy sector. It argues that energy efficiency and renewable energy are the main pillars of the energy transition, together with significant investment. Moreover, the report acknowledges the importance of regulatory framework and policy to give relevant stakeholders a guarantee that the transition will take place and to provide incentives that reflect climate change commitments. Similarly, the latest report on renewable energy by the European Environment Agency (EEA, 2017) considers renewable energy sources as the major contributor to the energy transition in Europe.

## 5. Case Study

European Union information will be used for our case study, and the rationale is two-fold. Firstly, the common energy policies developed at the EU level ensured us that action was being taken in the EU regarding energy transition. Secondly, the similar socio-economic situations resulting from the common membership enabled this research to isolate relevant factors more easily.

Firstly, common policies regarding energy transition adopted at the EU level make the EU a very relevant sample for this research. The key Directive regarding renewable energy was Directive 2009/28/EC, which promoted the use of energy from renewable sources. In December 2018, Directive 2018/2001 entered into force, revising Directive 2009/28/EC. While Directive 2009/28/EC encouraged Member States to achieve their national targets for renewable energy consumption by 2020, the Directive 2018/2001 set out national energy targets for 2030. Article 33 of the Directive provides for monitoring by the Commission (Directive 2018/2001). Such monitoring shall be based on Member States' reports or those of third countries or international organizations. Such provision encourages the gathering and sharing of data. Moreover, the EU publishes a renewable energy progress report every two years (European Commission, 2019a).

Secondly, EU membership conditions create relatively common socio-economic situations in every Member State, which facilitates the identification of relevant factors for energy transition. To look at a sample of relatively similar countries allows us to focus on relevant factors and exclude too important differences in the economic situation or the size of a country. The Copenhagen criteria define the conditions for accession. Candidate countries must have stable institutions guaranteeing democracy, the rule of law, human rights and respect for and protection of minorities. Moreover, they must have a functioning market economy and the capacity to cope with competition and market forces in the EU. Lastly, they must be able to take on and implement effectively the obligations of membership, including adherence to the aims of political, economic and monetary union (European Commission, 2016). The Copenhagen criteria filter out countries with too fragile markets or undemocratic conditions. This allows avoiding outliers in the sample and having more relevant and accurate results.

The EU also cooperates with several of its neighboring countries, and its policymaking thus extends to other European countries. Our research thus looks at EU Member States and neighboring European countries

## 5.1 Qualitative analysis

The various sources were chosen in the attempt to present a broad overview of the topic, regardless of disciplinary divisions. The articles, books and reports were gathered through research on google scholars and in libraries. The sources were chosen on the basis of their topics. Although through very different approaches, all treated of energy transition. For example, while some compare policy instruments regarding renewable energy (Fouquet & Johansson, 2008), others investigate education regarding renewable energy in schools (Korjakins, & al., 2018). Each source was listed in an excel sheet with 7 columns specifying the area of research, the scope of focus, the title, the authors, the publishing year, the publishing venues, and key findings. This allowed our research to have a clear overview of the sources covered. Afterwards, our research gathered factors explicitly or implicitly mentioned in each source. We define a factor as a measurable fact that - allegedly - contributes to a result, here the result being energy transition. For example, the number of laws regarding renewable energy, household income and the number of people who think positively about renewable energy are factors.

An important variety of factors was gathered varying significantly in character and complexity. However, many of them related to similar themes. It is been identified four recurrent themes in the literature. These themes were then considered as factors in categories for the rest of the analysis. The categories thus had to be specific enough to be different from each other, but broad enough to encompass a maximum amount of the factors. Four different themes were settled: market, government action, factors relating to the public, and technological advancement.

The market is identified as all the economic factors that can be connected to energy such as subsidies, taxes, production, prices and income level. Example of factors categorized in the market theme are investment in green energy supply or in renewable energy research (Midilli & al., 2006). Moreover, energy prices were also categorized under market, along with energy quantity demand and supply (Sung & Park, 2018; Energy Atlas, 2018). Energy is a commodity, so most authors assess of the transition by looking at how green energy is doing on the economic market.

Government action is here understood as a set of policies or actions both on national and supranational level in the relevant field. Here government is understood as all the public institutions of a state at the regional, national, supranational or international level. Examples of such factors are legislation and any regulatory framework or policy implemented (Zillman & al., 2018; Lund, 2007; Ruszel & al., 2017). Actions such as government subsidies or taxation were also considered government actions even though these were also categorized into the market theme (Lund, 2007; Bjørner & Jensen, 2002). As mentioned, the four themes overlap. Factors such as government investments are government actions that also concern the economic perspective of energy transition. Hence, these are counted in both themes as they reflect both the influence of government actions and of the market.

The theme public is understood as all factors that concern and influence the opinion of citizens. Example of factors categorized into public are public awareness, the level of environmental concern, education (Zaharo & Esa, 2016; Tsagarakis & al., 2018). The public theme is feasible to the research of data as many articles used surveys to assess public opinion over a matter or even conducted research to determine what influences people to adopt renewable energy. Moreover, the existence of the EU barometer made the public theme even more feasible regarding a research on the EU.

Technological advancements are understood as the generation of information or the discovery of knowledge that advances the understanding and development of technology in the field of energy and reduces energy costs. Examples of factors categorized into this theme are development of new forms of energy sources, the development of energy infrastructure (Midilli & al., 2006; Sattich, 2014; Ruszel & al., 2017). Moreover, energy efficiency was also considered in this theme (Rosenow & Kern, 2017).

These four themes give indication to what the literature considers relevant to green energy transition. These themes are mentioned in a high number of articles and hence are considered more relevant. Moreover, these themes were used a structural framework to search for data. Some other factors such as geography are considered relevant to energy transition by the academic discussion but are had less articles mentioning them and hence were not included as a fifth theme.

## 5.2 Quantitative analysis

The data used for the empirical and statistical analysis process were a subset of the bigger collection of datasets gathered after identifying the five themes of factors in the qualitative analysis. The inclusion of a given dataset within the calculations made depended on a few constraints. Firstly, the time frame of the dataset must be representative. This means that it must provide the same type of data for most of the EU member states, and a span of a reasonable time that would allow us to study the patterns of such indicator accurately. Secondly, the dataset should contain minimal missing values since this affects the results of statistical tests held on the data. Afterwards, the data collected would go through a cleaning process, making sure the values on which tests are computed are not biased because of a certain timeframe or because of the presence of an outlier value which pushes the mean of the sample distribution to a false value. To avoid this phenomenon, outliers are detected manually and removed before the computation of any statistics. This process is run for every dataset that is considered valuable and chosen as an indicator for one of the main themes. The paradigm we used for finding publicly accessible data related to the qualitative analysis results was search the most popular dataset search engines. The websites used to search for relevant datasets were Google Dataset Search, Kaggle, World Bank and EU Open Data Portal. The challenge following the collection of data from different sources is in the format; the aggregated data needs cleaning, as to be meaningful and match formats and types to allow operations on it.

The datasets which have a suitable time frame, include all the major EU-countries and do not have too many missing observations are:

R1, dependent variable: Renewable energy consumption in % of total final energy consumption (World Bank, Sustainable Energy for All (SE4ALL), 2019).

R2, dependent variable.: Renewable electricity consumption in % of total electricity consumption (World Bank, IEA Statistics, 2018).

A: Research and development budget (OECD, IEA, 2017).

B: Electric power consumption (kWh per capita) (World Bank, IEA Statistics, 2014).

C: GDP per capita (current US$) (World Bank national accounts data and OECD National Accounts data files, 2019)

D: Median and median income per household type (Eurostats, 2019).

E: Climate change law from the world database (Grantham Research Institute on Climate Change and the Environment and Sabin Center for Climate Change Law, 2019).

This results in a multivariate time series with 3372 data points of which 774 contained a missing value.

Additionally, in an attempt to identify patterns in the sample countries, the visualization program Tableau is used. This enables to visualize which countries are well on in the transition, and the ones that are lacking behind. The results are presented in the next section.

### 5.3 Results

This section presents the results of the visualization in Tableau, the pre-processed data is plotted to reveal underlying correlations between different factors and the indicators of green energy transition. The later was taken in this research to be the percentage of green energy production and green energy consumption over total energy production and consumption respectively. The statistical methods run on the data is calculating Pearson's correlation coefficient for combinations of all the pre-processed data which represented the factors. The results of this statistic revealed an initial result that there is no linear relationship between any pairs of factors, supported by the average correlation coefficient of approximately 0.084, clearly a small value. Although a maximum coefficient occurred at approximately 0.95, this is the correlation between the Co2 emissions dataset and the gas emissions dataset, which both represent the same factor:

serving as an indicator of green energy transition. Following the initial tests, the same data was used in visualization tools to check for possible empirical patterns and compare them to the statistical results.

Firstly, by mapping the countries in the data analysis software Tableau, clusters of poor performing countries and good performing countries are separated. Performance is measured with the average renewable electricity output. Among other important observations, one can identify clusters of well performing countries in Scandinavia and the Balkan States. To the contrary, Eastern and Central Europe and the United Kingdom (UK) perform overall worse than the average. Thanks to the mapping, one could additionally observe a substantial difference in average renewable energy output for Latvia and Estonia. This is puzzling as those are two seemingly similar countries regarding their politics, history and institutions. This puzzle is addressed in the results section.

Secondly, to identify patterns that indicate a causal relationship between the average renewable electricity output and one of the other indicators, the best and worst performing countries are grouped. The outperforming group in terms of green energy transition consists of Island, Norway, Sweden, Latvia, Switzerland, Austria, Croatia, Bosnia Herzegovina, Montenegro and Albania. The worst performing group consists of Ireland, the UK, Netherlands, Belgium, Belarus, Poland, Czech Republic, Kosovo, Estonia and Ukraine. There are three underlying assumptions tested. Firstly, the good performing countries have on average a higher GDP per Capita. Secondly, the good performing countries have a higher number of legislations acts regarding green energy policies. Thirdly, the good performing countries have a higher share of Research and Development investments (R&D) per GDP.

In contrast to the assumptions, there are no convincing patterns visible. Regarding the first assumption, GDP per capita, the average of the further transitioned countries is 36.663 Euro, whereas the lower transitioned countries average only 29.291 Euro. This result confirms the first assumption, but only to a certain extent as this factor cannot be yet considered as significant. Regarding the second assumption, the visualization for the number of legislation acts shows that both country groups implement the same number of legislation acts corresponding to renewable energy. This finding suggests that there is no direct relation between this factor and the energy transition. Surprisingly, the worst performing country group averages with 18.5% compared to 13.1% a substantially higher share of R&D/GDP. This result suggests no direct influence of R&D on the renewable energy output. This further leads towards the idea that a process as complex as green energy transition is not only based on one factor, but rather on multiple factors like natural resources, politics, industrial strategy and the world energy prices. Explanations for this difference could be the sectoral composition of the individual countries, as certain sectors have a lower R&D level, or efforts of the worst country group to catch-up with the competing country group.

A third visualization plotted the countries in a histogram to identify the development of renewable energy share of the overall energy output in the given time horizon. The focus of this part is to isolate a group of countries that displays a steady and steep increase in the share of renewable energy output. By excluding all poor performing countries, good performing countries that did not show any steep increase and countries with unusual observations, three countries are left for further investigation. The best development, and therefore serving as a case example, is Denmark. Denmark steadily improved from 2002-2017 from 17.3% to 65.5%. Lower in absolute increase, but with a substantial

improvement in relative terms, the UK and Belgium are also a positive role models for other countries

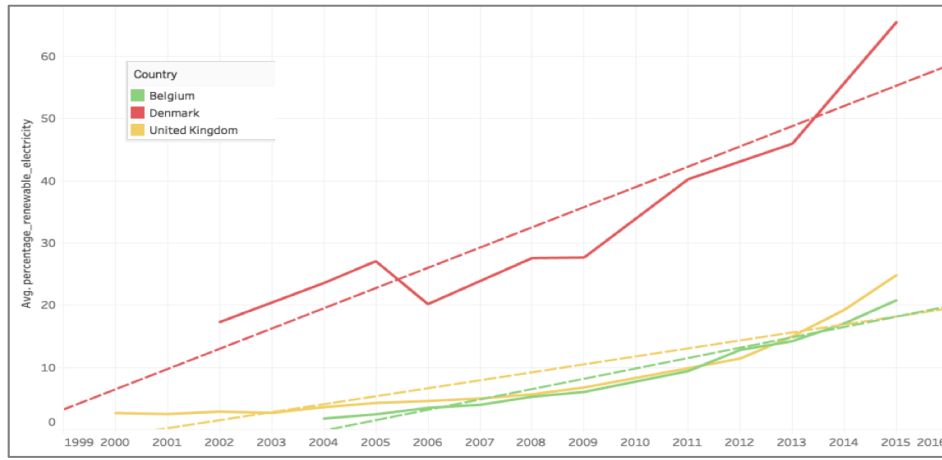

**Figure 1.** Best Improvement on average renewable electricity output. Belgium, Denmark and United Kingdom

Despite the fact that energy transition, climate change and sustainability are receiving increased attention in the world headlines, the academic literature analyzing energy transition specifically at the EU level is rather scarce. Our systematic review of papers evaluating the relevant factors behind energy transition identified only a few comprehensive studies on the topic, none explicitly focusing on providing a pan-European analysis on its specific drivers. Some reports have been conducted looking at energy transition in the EU specifically (European Commission, 2019b; Energy Atlas, 2018). However, these do not delve into the factors influencing the phenomenon but instead focus on the current state and trajectory of the transition. Sung & Park (2018) carried out a multidisciplinary analysis on the phenomenon looking at the different actors and conducting empirical analysis. However, they did not explicitly evaluate EU countries as such and did not directly look at technological advancement as a relevant factor.

The majority of studies look at energy transition by either discussing it as a subtopic to another phenomenon, or only focusing at one or a few of the factors possibly at stake. Among the different factors relevant to energy transition, policymaking appears to be the most studied sector regarding energy transition. However, even within this domain, papers taking a multi-disciplinary approach are scarce. The relevance of investment and economic development in energy transition cannot be ignored. Moreover, the public awareness of the topic possibly influences government and market actors. Technological advancement, specifically the development of renewable energy sources and energy efficiency, can significantly contribute to energy transition. In reality, none of the factors functions completely independently and any successful energy transition strategy will need to acknowledge the power of each factor as well as their inter-connection.

The literature review also showed that no clear consensus exists on the definition of energy transition. However, specifying the boundaries and dimensions of energy transition is crucial. Hence, priorities need to be set in order to address the issue. Meadowcraft (2009) rightly states that what sort of transition is embarked upon, and thus which approaches, and technologies are favored and supported, has important consequences for society and cannot be left to take its own course.

No paper evaluating energy transition in the EU took a comprehensive and broad approach, including as many relevant factors as possible. Moreover, none conducted such work through a qualitative and quantitative analysis. Although there are evaluations of EU policies on energy transition, none considers the aggregate benefits and inconvenient of such policies for the whole union. This is a relevant gap in research, which is important to be addressed in order to provide aid to EU policymakers. The importance of the issue, as demonstrated above, requires immediate but wise action, and hence any measures should be based on deep knowledge of the different issues underlying energy transition.

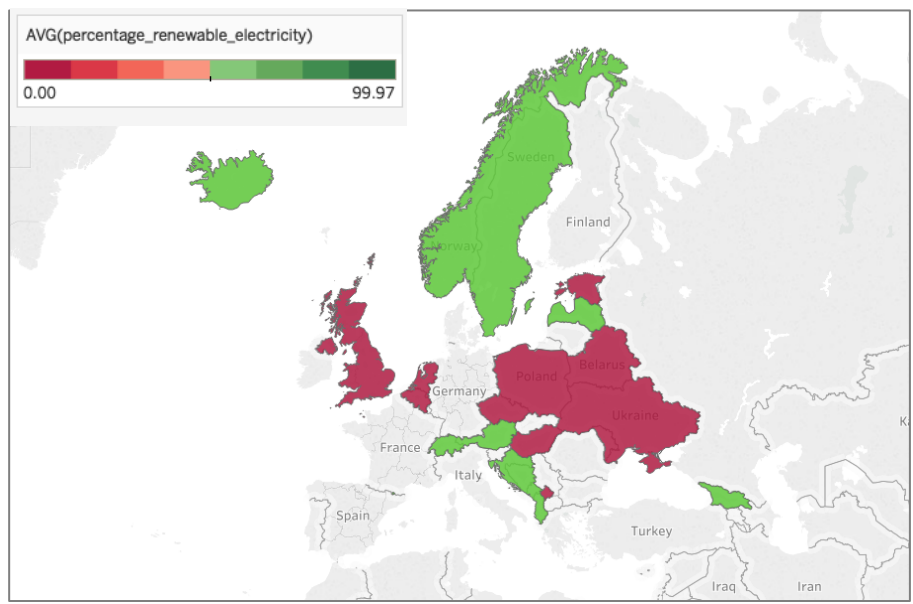

**Figure 2.** Overview High vs Low Percentage renewable Electricity Output

The EU has already come a long way with its ambitious goal of reducing greenhouse gas emissions, which is ultimately the main goal of energy transition. However, recent studies have argued that this reduction might not be caused by a successful ongoing energy transition but by a transfer of emissions to developing countries (Intergovernmental Panel on Climate Change, 2014; Peters, Minx, Webber & Edenhofer, 2011). They suggest that the stabilization of emissions in developed countries, such as the EU Member States, is partially due to growing imports from developing countries. Since lower environmental standards apply in these countries, Europe might not be decreasing its emissions but relocating them. Ignoring the impact of international trade could result in a misleading analysis of the underlying forces of regional emission trends and energy transition policies.

Lastly, as argued by other authors (Sung & Park, 2018), the inadequacy of accessible data limits our ability to make firm conclusions on how certain factors influence the current energy transition. Moreover, studies voiced that the lack of accessible, relevant data on energy transition in the EU hinders research on the topic and development in the sector of renewable energy (GGKP, 2017; IEA & the World Bank, 2017). Although national data on energy transition must exist, no dataset including every country of the union is made available by the later. The EU Eurostat service offers different datasets, but none of the accessible data was relevant to our research. Further research is thus called for in not only identifying how different factors contribute to energy transition but also in improving the quality and availability of relevant data.

## 6. Conclusions.

This research asked what energy transition is and what are the different factors responsible for energy transition. Through a systematic literature review, our research identified four recurrent themes in the academic literature and official reports regarding energy transition, namely, market, government action, public and technological advancements. These themes were used as factors in the rest of the research. Data sets were collected and the relevant and useable one was analyzed through the visualization data analysis software Tableau. Preliminary results shown that GDP per capita and household net income are potential candidates to predict energy transition best. Moreover, a look into the Balkan State energy transition highlighted the importance of policies in creative incentive for energy transition.

However, the key findings of our research are that the definition of energy transition is not understood in a homogeneous way in the academic literature. This leads to different definitions and different approaches regarding energy transition. Moreover, multi-disciplinary work on the phenomenon is not numerous. All of this hinders our understanding of what is an energy transition. It is expected that the energy transition debate will continue, hopefully in a widened sense, taking into account all the critical factors with an interdisciplinary approach.

A second important mark of our research is the poor accessible, workable and relevant data on the factors that drive green energy transition in a comprehensive manner. This research did not look into factors such as geography and other possibly relevant factors, such as whether the outsourcing of pollution in Europe by increased import from developing countries should be taken into account when assessing a country's energy transition. Future research should look into those factors while still including socio-economic factors, taking a broad, all-encompassing perspective on energy transition.

If the EU adopts a truly multi-disciplinary approach to energy transition, making the most of the knowledge and data gathered on Member States as well as internationally, the EU can lead in the transition towards a more sustainable and clean energy system.

**Acknowledgements:** The authors would like to thank the Maastricht University Honours + students' class 18-19 who contributed on writing this report: Léïa Bonjwean, Liselotte Grönlund, Samuel Kopp, Marwan Mansour, Simon Stachelscheid, and Michele Monaco.